\documentstyle[12pt]{article}
\def\x{\stackrel{\otimes}{,}}

\def\a{\begin{eqnarray}}
\def\b{\end{eqnarray}}
\def\0{\nonumber}
\input amssym.def
\font\teneusm=eusm10                    % loads Euler script medium weight
\font\seveneusm=eusm7                   % (Euler frak (eufm) already loaded
\font\fiveeusm=eusm5                    % by amssym.def)
\newfam\eusmfam
\textfont\eusmfam=\teneusm
\scriptfont\eusmfam=\seveneusm
\scriptscriptfont\eusmfam=\fiveeusm

\def\sG{sine--Gordon}
\def\shG{sinh--Gordon}

\def\dg{ dressing group}

\renewcommand{\theequation}{\thesection.\arabic{equation}}

\newlength{\extraspace}
\newlength{\extraspaces}
\newcounter{dummy}

\newcommand{\ai}{
\addtocounter{equation}{1}
\setcounter{dummy}{\value{equation}}
\setcounter{equation}{0}
\renewcommand{\theequation}{\thesection.\arabic{dummy}\alph{equation}}
\begin{eqnarray}
\addtolength{\abovedisplayskip}{\extraspaces}
\addtolength{\belowdisplayskip}{\extraspaces}
\addtolength{\abovedisplayshortskip}{\extraspace}
\addtolength{\belowdisplayshortskip}{\extraspace}}
\newcommand{\bj}{
\end{eqnarray}
\setcounter{equation}{\value{dummy}}
\renewcommand{\theequation}{\thesection.\arabic{equation}}}

\input amssym
\newcommand{\be}{\begin{equation}}
\newcommand{\ee}{\end{equation}}
\newcommand{\ba}{\begin{eqnarray}}
\newcommand{\ea}{\end{eqnarray}}
\newcommand{\ban}{\begin{eqnarray*}}
\newcommand{\ean}{\end{eqnarray*}}
\newcommand{\brr}{\begin{array}}
\newcommand{\err}{\end{array}}
\newcommand{\bc}{\begin{center}}
\newcommand{\ec}{\end{center}}

\def\E{{\cal E}}

\def\F{{\cal F}}

\def\T{{\cal T}}

\def\l{\lambda}
\def\z{\zeta}

\def\be{\beta}

\def\ep{\epsilon}

\def\var{\varphi}
\newcommand{\bea}{\begin{eqnarray}}
\newcommand{\eea}{\end{eqnarray}}
\newcommand{\bean}{\begin{eqnarray*}}
\newcommand{\eean}{\end{eqnarray*}}
\newcommand{\ZZ}{\Bbb Z}

\newcommand{\del}{\partial}

\newcommand{\dpiu}{\partial_+}
\newcommand{\dmeno}{\partial_-}
\begin{document}
 
\begin{titlepage}
 
\begin{flushright}
CBPF.NF.076/95 \\
 hep--th/9510174
\end{flushright}
\vskip0.5cm
\centerline{\LARGE   A Note on the Symplectic Structure on the Dressing }
\centerline{\LARGE    Group  in the sinh--Gordon Model }
\vskip1.5cm
\centerline{\large   Guillermo Cuba
\footnote{E--mail address gcubac@cbpfsu1.cat.cbpf.br}     
 and Roman Paunov
\footnote{E--mail address paunov@cbpfsu1.cat.cbpf.br}}
\centerline{Centro Brasileiro de Pesquisas Fisicas }
\centerline{Rua Dr. Xavier Sigaud 150, Rio de Janeiro, Brazil}

\vskip5cm
\abstract{ We analyze the symplectic structure on the dressing group in the \shG\, model by calculating explicitly the Poisson bracket $\{g\x g\}$
where $g$ is the \dg\, element which creates a generic one soliton solution from the vacuum. Our result is that this bracket  does not coincide with the Semenov--Tian--Shansky one. The last induces a Lie--Poisson structure on the 
\dg . To get the bracket obtained by us from the 
Semenov--Tian--Shansky bracket we apply the formalism of the constrained 
Hamiltonian systems. The constraints on the \dg\, appear since the element
which generates one solitons from the vacuum has a specific form.}
 
\end{titlepage}

\section{ Introduction}

\setcounter{equation}{0}
\setcounter{footnote}{0}

The solitons are particle like solutions of the integrable non--linear equations \cite{Fad},\cite{Nov}. They describe elastic collision of solitary waves. After the interaction, the outgoing waves propagate with the same rapidities as the ingoing ones but with changed phases. The \sG\, theory is an example of an integrable model both at the classical and at the quantum level. The quantum scattering matrix
was constructed in \cite{Zam} by using bootstrap methods. It is well known \cite{ZZ} that the quantum \sG\, model for certain values of the coupling constant is a massive integrable perturbation of the minimal conformal models in two dimensions.

In \cite{dress} the \dg\, symmetry was proposed as an alternative approach to solve classical integrable models. It was also argued that the \dg\, is a semiclassical limit of the quantum group symmetry of a quantum integrable model. The action of the \dg\, is realized via gauge transformations 
which act on the components of the Lax connection. A non--trivial Poisson bracket,  known as 
Semenov--Tian--Shansky bracket was introduced \cite{Sem} in order to ensure the covariance of the 
Poisson brackets on the phase space under the \dg\, action. The \dg\, together with the Semenov--Tian--Shansky bracket becomes a Lie--Poisson group.

In \cite{BaBe} the  \dg\, elements which generate $N$--solitons from the vacuum
in the \shG\, model and in its conformally invariant extension \cite{BB} are constructed explicitly.   
 It was also shown that there is a relation between the \dg\, and the vertex operator construction of the soliton solutions  \cite{BaBe}, \cite{Kyoto}.

In the present note we calculate the Poisson bracket $\{ g \x g\}$ of the dressing group element $g$ which generates an arbitrary one soliton solution in the \shG\, model from the vacuum. We use the fact that the phase space of the one solitons is two dimensional \cite{Fad}, \cite{NBB}. Surprisingly, 
the Poisson bracket found by us is {\it not} identical to the Semenov--Tian--Shansky expression which is proportional to $[r, g\otimes g]$ where $r$ is the classical $r$--matrix \cite{Fad}. 
Instead of the later we obtain the expressions (\ref{fin}), (\ref{risultato}). 
It is known \cite{BaBe} that the \dg\, element which produces one solitons from the vacuum has a special form. To reproduce it one has to impose constraints
on the \dg . We apply the formalism of the constrained Hamiltonian systems
\cite{vincoli} to show that our result for the bracket $\{g \x g\}$ follows from the Semenov--Tian--Shansky bracket after a suitable reduction on the
\dg .

We outline the content of the paper. Sec. 2 is devoted to the one soliton solutions of the \shG\, equation and to the construction of the \dg\, element which produces the one solitons from the vacuum. In Sec. 3 we present our calculation of the Poisson bracket of the \dg\, element and express it with the 
use of the classical $r$--matrix. In Sec. 4 we apply the general principles of the constrained dynamics in order to establish a relation between  our bracket
and the Semenov--Tian--Shansky bracket.

\section{The \shG\, equation, the one soliton solutions and 
the dressing group.}

\setcounter{equation}{0}
\setcounter{footnote}{0}
 
In this chapter we briefly review some basic facts concerning the 
\shG\, model, its one soliton solutions and the dressing 
group \cite{Fad}, \cite{Nov}, \cite{dress},  \cite{BaBe}. We start by recalling the \shG\, equation in two dimensions
\a
\dpiu \dmeno \var &=&2 m^2 sh 2\var ~~~~~~ \del_{\pm}=
\frac{\del}{\del x^{\pm}}\0\\
x^{\pm}&=& x\pm t \label{shG}
\b
It is clear that it has a vacuum solution $\var=0$. The eq. (\ref{shG})
is equivalent to the zero curvature condition 
$F_{+-}=\del_+ A_- -\del_-A_+ +[A_+, A_-]=0$ of the connection
\ai
A_{\pm}&=& \pm \del_{\pm} \Phi + m e^{\pm ad \Phi} \E_{\pm}
\label{con}\\
\E_{\pm}&=& \l^{\pm 1} \left( E^++ E^-\right) ~~~~~~
\Phi=\frac{1}{2} \var H
\label{Heisenberg}
\bj
where $E^{\pm}$ and $H$ are the generators of the $sl(2)$ Lie algebra
\a
[H,E^{\pm}]=\pm 2 E^{\pm}~~~~~~ [E^+,E^-]=H \label{sldue}
\b
and $\l$ is the spectral parameter.
The components of the Lax connection (\ref{con}) belong to the loop algebra $\tilde{sl}(2)$. In the {\it principal gradation} the last is generated by  the elements $E^{\pm}_{n}=\l^n E^{\pm}, \,\,\,\, n\in 2\ZZ$ $+1$ and $H_{n}=\l^n H, \,\,\,\, n\in 2\ZZ$..
The commutation relations are the following
\a
[H_k, E^{\pm}_l]=\pm 2 E^{\pm}_{k+l}~~~~~~ 
[E^+_k,E^-_l]=H_{k+l}\label{loop}
\b
The flatness of the connection (\ref{con}) implies that there 
exists a solution of the linear system
\a
\left( \del_{\pm} + A_{\pm} \right) T(x^+,x^-,\l)=0
\label{ls}
\b
which together with the initial condition $T(0,0,\l)=1$ is known in the literature as a
normalized transport matrix.

The canonical symplectic structure $\{\del_t \var (x,t), \var (y, t)\}=
\delta (x-y)$ can be equivalently written in the form
\ai 
\{A(x,t) \x A(y,t)\} &=& \frac{1}{4} [r, A(x,t)\otimes 1+
1\otimes A(y,t)]\delta (x-y) \label{basic}\\
r&=& -\frac{\l^2+\zeta^2}{\l^2-\zeta^2} H\otimes H\0\\
-&4& \frac{\l \zeta}
{\l^2-\zeta^2} \left( E^+\otimes E^- +E^-\otimes E^+\right)
\label{rmat}
\bj
where $A=A_++A_-$ is the spatial component of the Lax connection;
$\l$ and $\z$ are the spectral parameters corresponding the left and the right tensor factors respectively.

To introduce the one soliton solutions we consider the variable $\ep^+(x^+,x^-)=\ep^+(x)$ whose dependence on the light cone variables is dictated by the relation
\a
\frac{ \ep^+(x)+\mu}{\ep^+(x)-\mu}&=& a 
\exp \{ 2m(\mu x^++ \frac{x^-}{\mu}) \}\0\\
a&=&\frac{ \ep^++\mu}{\ep^+-\mu},\,\,\,\,\,\, \ep^+=\ep^+(0,0)
\label{ev}
\b
The \shG\, field is expressed in terms of $\ep^+$ and the soliton rapidity $\mu$ as follows
\a
e^{-\var(x)}&=& -\frac{\ep^+(x)}{\mu}
\label{campo}
\b
We shall also need the variable $\ep^-(x)$ related to $\ep^+(x)$ and $\mu$ by 
$\ep^+(x)\ep^-(x)=\mu^2$ or equivalently
\a
\frac{ \ep^+(x)+\mu}{\ep^+(x)-\mu}&=-&\frac{ \ep^-(x)+\mu}{\ep^-(x)-\mu}
\label{id}
\b

 In order to construct a solution of the linear problem (\ref{ls}) one observes that 
(\ref{ev}) is equivalent to
\a
\psi_{\pm}(x,\mu)&=& \pm a \psi_{\pm}(x,-\mu)
\label{symm}
\b
where
\a
\psi_{\pm} (x,\l)&=& \left(\ep^{\pm}(x)+\l\right)  
e^{-m ( \l x^+ +  \frac{x^-}{\l})}
\label{jap}
\b

It was shown in \cite{Date} that the matrix 
\a
 {\bf \T}&=&e^{\Phi(x)}\left (\begin{array}{cc}  \psi_+(x,\l)& \psi_+(x,-\l)\\ 
\psi_-(x,\l)&  -\psi_-(x,-\l)
\end{array}\right )\label{psi}
\b
satisfies the linear system (\ref{ls}) with $\var$ being  the one soliton solution ({\ref{campo}).
Due to (\ref{symm}), the matrix $\T$ as a function on the spectral parameter $\l$ is degenerated at the points $\l=\pm \mu$. A direct calculation shows that 
\a
\det \T &=& 2 (\l^2-\mu^2)
\label{det}
\b
The normalized transport matrix is $T(x,\l)=\T(x,\l) \cdot \T^{-1}(0,\l)$.

Starting from (\ref{psi}) one can construct algebraically solutions of the dressing problem. 
First we recall that the \dg\, element which relates the vacuum to an one soliton solution is 
introduced by the equation
\a
T(x,\l)&=& g(x,\l) \cdot T_0(x,\l) \cdot g^{-1}(0,\l)
\label{vestimento}
\b
where $T_0(x,\l)=\exp \{ -m(x^+\E_++x^-\E_-)\}$ is the vacuum solution transport matrix. The last is expressed as $T_0(x,\l)=\T_0(x,\l)
\T^{-1}_0(0,\l)$; $\T_0(x,\l)$ has the same form as (\ref{psi})
but with $\Phi =0$ and $\psi_+(x,\l)=\psi_-(x,\l)= e^{-m ( \l x^+ +  \frac{x^-}{\l})}$.
The element 
$g(x,\l)$ is expressed in terms of the non--normalized transport matrices $\T$ and $\T_0$ as follows
\ai
g(x,\l)&=& \T(x,\l)\cdot \T_0^{-1}(x,\l)\cdot S(\l) \label{furbo}\\
S(\l)&=&\left (\begin{array}{cc}  a(\l)& b(\l)\\ 
b(\l)&  a(\l)
\end{array}\right )\label{const}\\
\det S(\l)&=& \frac{1}{\mu^2-\l^2}\label{sl}\\
a(\l)&=&a(-\l)\,\,\,\,\,\,\, b(\l)=-b(-\l) \label{principale}
\bj
The form of the $x^{\pm}$--independent matrix $S(\l)$ (\ref{const}) is fixed by the requirement that it should commute   with the matrix $T_0(x,\l)$. The equation (\ref{sl}) guarantees that 
the dressing group element $g(x,\l)$ has a unit determinat; the condition (\ref{principale}) reflects the fact that it is represented in the principal gradation.
Setting $a(\l)+b(\l)=\l-\mu$ and $a(\l)-b(\l)=-\l-\mu$ one recovers the solution constructed in 
\cite{BaBe} 
\a
g(x,\l)&=&\frac{e^{\Phi(x)}}{2 (\l-\mu)}
\left (\begin{array}{cc} 
 \l+\ep^+(x)& \l+\ep^+(x) \\ 
\l+\ep^-(x)&  \l+\ep^-(x)
\end{array}\right )+\0\\
&+&\frac{e^{\Phi(x)}}{2 (\l+\mu)}
\left (\begin{array}{cc}  \l-\ep^+(x)& -\l+\ep^+(x)) \\ 
-\l+\ep^-(x)& \l-\ep^-(x)
\end{array}\right )\label{mutandina}
\b
We  note that the above expression is not the unique which satisfies the following requirement: the matrix elements of $g(x,\l)$ are meromorphic functions on the Riemann sphere with only simple poles at the points $\l=\pm \mu$. There exist four solutions: $a(\l)+b(\l)=(-)^p (\l-(-)^k\mu)$, 
$a(\l)-b(\l)=(-)^{p-1}(\l+(-)^k\mu)$ where $p, \, \, k=0,\,\, 1$. These solutions have the following assymptotic  behaviour
\a
g(x,\l)&=& (-)^p e^{\Phi(x)} +O(\l^{-1}) \,\,\, \l \rightarrow \infty \0\\
g(x,\l)&=& (-)^{p+k} e^{-\Phi(x)} +O(\l) \,\,\, \l \rightarrow 0
\label{limite}
\b
The solution (\ref{mutandina}) is good since it turns to $e^{\Phi}$ for $\l\rightarrow \infty$ and
to $e^{-\Phi}$ when $\l \rightarrow 0$ \cite{dress}. More than that it permits to make a relation with the vertex operators.

\section{Derivation of the Poisson bracket of the \dg\, elements }

\setcounter{equation}{0}
\setcounter{footnote}{0}
 
In this paper we restrict ourselves to calculate the bracket $\{g\x g\}$ for (\ref{mutandina}) only. The other choices for $g$ mentioned above will be analyzed in \cite{Gui}. We shall use the coordinates $\ep^+$ and $\mu$ to parametrize the phase space of the one solitons. It is clear that
\a
\{g(\l)\x  g(\zeta)\}&=&
 \left(\frac{ {\partial g(\l) }}    { {\partial \ep^+} }\otimes  \frac{ {\partial g(\zeta)} }{ {\partial \mu} }-  \frac{\partial g(\l)}{\partial \mu}\otimes  \frac{\partial g(\zeta)}{\partial \ep^+}\right) \{ \ep^+ \, , \, \mu \}
\label{PB}
\b
We recall that (\ref{mutandina}) contains dependence on the variable $\ep^-$ but the last is a function on $\ep^+$ and $\mu$ according to (\ref{id}). To get an explicit expression for 
(\ref{PB}) we first obtain
\a
\frac{{ \del g(\l)}}{{\del \ep^+}}g^{-1}(\l)&=& 
-\frac{1}{2\ep^+(\l^2-\mu^2)}
\left(  (\l^2+\mu^2)H+ 2\l\mu (E^+-E^-) \right)\0\\
\frac{ {\del g(\l)}}{{\del \mu}} g^{-1}(\l)&=&
\left( \frac{\l^2+\mu^2}{2\mu(\l^2-\mu^2)}+
\frac{((\ep^+)^2-\mu^2)\l^2}{\ep^+(\l^2-\mu^2)^2} \right)H-\0\\
&& \hskip -3.5cm -\frac{\l\mu(\l^2-(\ep^+)^2)}{\ep^+(\l^2-\mu^2)^2}E^+-
\frac{\ep^+}{\mu}\left( \frac{\l}{\l^2-\mu^2}+2 \frac{\l\ep^-}{\mu(\l^2-(\ep^-)^2)}\right)
\frac{\l^2-(\ep^-)^2}{\l^2-\mu^2}E^-
\label{derivate}
\b

In what follows it will be convenient to introduce the following elements of the loop algebra
$\tilde{sl}(2)$
\a
X^0(\l)&=& \frac{1}{\l^2-\mu^2}\left( (\l^2+\mu^2) H+ 2\l\mu(E^+-E^-)\right)\0\\
X^{\pm}(\l)&=& \pm \frac{\l \mu}{\l^2-\mu^2}\left( H +\left( \frac{\l}{\mu}\right)^{\pm 1} E^+-
\left( \frac{\mu}{\l}\right)^{\pm 1} E^- \right)
\label{base}
\b
where we have omited the dependence on $\mu$ since we are going to keep the value of the rapidity fixed. The generators (\ref{base})
 are eigenvectors of the adjoint action of the element (\ref{mutandina})
\a
g(\l) X^0(\l) g^{-1}(\l)&=& X^0(\l)\0\\
g(\l) X^{\pm}(\l) g^{-1}(\l)&=& e^{ \pm \var} X^{\pm}(\l)
\label{aggiunta}
\b
We note that the expressions (\ref{base}) provide two embeddings  (one can consider the generators (\ref{base}) as expansions on positive or negative powers on $\l$)   of the $sl(2)$
Lie algebra (\ref{sldue}) in the loop algebra $\hat{sl} (2)$ which are given by
\a
H &\rightarrow & X^0(\l) \0\\
E^{\pm} &\rightarrow & X^{\pm}(\l)
\label{sldueaff}
\b
The fundamental two dimensional representation of the Lie algebra $sl(2)$ in the realization (\ref{base}) is spanned on the $\l$--depending vectors
\a
e_+(\l)= \left (\begin{array}{c} 
 1 \\ 
-\frac{\mu}{\l}
\end{array}\right ),~~~~~~~~~~
e_-(\l)=\left (\begin{array}{c} 
-\frac{\mu}{\l}\\1
\end{array}\right )\label{fund}
\b
We also introduce the  representation: $\tilde{X}^0=(X^0)^t,\,\,\,
\tilde{X}^{\pm}=(X^{\mp})^t$ where the upper index $t$ stands for the matrix transposition. The corresponding basis  dual to (\ref{fund}) is given by
\a
\tilde{e}_+(\l)= \frac{\l^2}{\l^2-\mu^2}\left (\begin{array}{c} 
 1 \\ 
\frac{\mu}{\l}\end{array}\right ), ~~~~~~~~~~
\tilde{e}_-(\l)= \frac{\l^2}{\l^2-\mu^2}\left (\begin{array}{c} 
\frac{\mu}{\l}\\1
\end{array}\right )\label{cofund}
\b
Obviously $(\tilde{e}_a,\, e_b)= \tilde{e}_a^t \cdot e_b=\delta_{ab}$ for
$a,\, b=\pm$; $X^+(\l)e_+(\l)=\tilde{X}^+(\l)\tilde{e}_+(\l)=0$

In terms of (\ref{base}) the derivatives (\ref{derivate}) are expressed as follows
\a
\frac{{ \del g(\l)}}{{\del \ep^+}}g^{-1}(\l)&=& -\frac{1}{2\ep^+}X^0(\l)\0\\
\frac{ \del g(\l)}{\del \mu} g^{-1}(\l)&=&\frac{1}{2\mu}X^0(\l)-\0\\
&-&\frac{\l}{\mu(\l^2-\mu^2)}\left( (\mu+\ep^-)X^+(\l)+(\mu+\ep^+)X^-(\l)\right)
\label{buono}
\b
Substituing back the above expression into (\ref{PB}) we arrive at the expression 
\a
\{g(\l) \x g(\z)\}\cdot g^{-1}(\l) \otimes g^{-1}(\z)&=&\0\\
&&\hskip -7cm - \frac{1}{2} \{ \ln \ep^+ ,\mu\}
\frac{\l}{\mu (\l^2-\mu^2)}
\left( (\mu+\ep^-)X^+(\l)+(\mu+\ep^+)X^-(\l)\right)\otimes X^0(\z)+\0\\
&&\hskip -7cm  +\frac{1}{2}\{\ln \ep^+ ,\mu\}
\frac{\z}{\mu (\z^2-\mu^2)} X^0(\l)\otimes
\left( (\mu+\ep^-)X^+(\z)+(\mu+\ep^+)X^-(\z)\right)
\label{fin}
\b

In the basis (\ref{base}) the $r$--matrix (\ref{rmat}) 
reads
\a
r&=& -2 \frac{  \l \mu }{\l^2-\mu^2}
\left( X^+(\l)-X^-(\l) \right) \otimes X^0(\z)+\0\\
&+& 2 \frac{ \z \mu }{ \z^2-\mu^2} X^0(\l)\otimes
\left( X^+(\z)-X^-(\z) \right)+ \ldots
\label{ere}
\b
where we have omited terms $X^0\otimes X^0$ and $X^{\pm}\otimes X^{\mp}$ since they do not contribute to the commutator
$[r, g\otimes g]$. 

Taking into account (\ref{campo}), (\ref{aggiunta}) and (\ref{ere}) we 
get the following expression for the bracket (\ref{fin})
\a
\hskip - 2cm \{g(\l)\x g(\z)\}&=&\frac{\left( Ad^{-1} g(\l)\otimes Ad^{-1} g(\z)
- Ad g(\l)\otimes Ad g(\z)\right)}{8}\cdot [r\, , g(\l)\otimes g(\z)]
\label{risultato}
\b
In the derivation of the above equation we have also used the bracket
\a
\{\ep^+, \mu \}&=&-\frac{1}{2}\left( (\ep^+)^2-\mu^2\right)
\label{body}
\b
which was calculated in \cite{Fad}, \cite{NBB}.

The bracket (\ref{risultato}) obviously differs from the Semenov--Tian--Shansky 
expression which reads
\a
\{g(\l)\x g(\z)\}_{STS}&=&\frac{1}{4}[  r\, , g(\l)\otimes g(\z)]
\label{STS}
\b
We point out that (\ref{STS}) was derived in \cite{dress}  \cite{Sem} from the 
requirement that the \dg\, action is a Lie--Poisson action, i. e. the  Poisson brackets of the model transform covariantly under the dressing transformations.
On the other hand, the result (\ref{risultato}) is a consequence of the 
bracket (\ref{body}) which reflects the fact that the one--solitons can be represented as an one--body relativistic problem \cite{NBB}. This observation
suggests that (\ref{risultato}) should be obtained from (\ref{STS}) by applying the theory of the constrained Hamiltonian systems. This will be done in the next section.

\section{The Relation between the two brackets}

\setcounter{equation}{0}
\setcounter{footnote}{0}

A generic \dg\, element $f(\l)\in \tilde{SL}(2)$ is represented by a $2\times 2$ matrix with unit determinant, the elements of which $f_{ij}(\l),\,\,\, i,j=1,2$
are functions on $\l$.  The element (\ref{mutandina}) has  a special form, and due to that we have to 
impose  constraints on the \dg\, in order to obtain it. To do that we first introduce the Gauss--like decomposition
\a
f(\l)&=&e^{a_-(\l)X^-(\l)}e^{\ln \kappa(\l)X^0(\l)} e^{a_+(\l) X^+(\l)}
\label{Gauss}
\b
 We note that, due to (\ref{base}) for $\l \rightarrow \infty$ one ends up with the  decomposition of the Lie group $SL(2)$:
$f=e^{a_-(\infty)E^-} e^{\ln \kappa(\infty) H} e^{a_+(\infty) E^+}$ while in the limit $\l \rightarrow 0$ the factorization $f=e^{a_-(0)E^+} 
e^{\ln \kappa (0)} e^{a_+(0) E^-}$ is valid. Therefore, in these limits
we obtain  two versions of the Gauss decomposition of the classical Lie group $SL(2)$. We stress that (\ref{Gauss})
has nothing to do with the Gauss decomposition of the loop group.

Taking into account (\ref{sldueaff}), (\ref{fund}) and (\ref{cofund})
we obtain
\ai
\kappa(\l)&=&\frac{\l^2}{\l^2-\mu^2}\left(f_{11}(\l)-\frac{\mu}{\l}
(f_{12}(\l)-f_{21}(\l))-\frac{\mu^2}{\l^2}f_{21}(\l)\right)\\
\label{kap}
a_+(\l)&=&\frac{f_{12}(\l)-\frac{\mu}{\l}(f_{11}(\l)-f_{22}(\l))-\frac{\mu^2}{\l^2}f_{21}(\l)}
{f_{11}(\l)-\frac{\mu}{\l}
(f_{12}(\l)-f_{21}(\l))-\frac{\mu^2}{\l^2}f_{21}(\l)}\\
\label{apiu}
a_-(\l)&=&\frac{f_{21}(\l)+\frac{\mu}{\l}(f_{11}(\l)-f_{22}(\l))-\frac{\mu^2}{\l^2}f_{12}(\l)}
{f_{11}(\l)-\frac{\mu}{\l}
(f_{12}(\l)-f_{21}(\l))-\frac{\mu^2}{\l^2}f_{21}(\l)}
\label{amen}
\bj
The inverse map is given by
\a
f_{ij}(\l)&=&(-)^i\frac{\l\mu}{\l^2-\mu^2}\left(\frac{\l}{\mu}\right)^{-i-j+3}\0\\
&\times & \left( (a_+(\l) +(\frac{\l}{\mu})^{-2j+3})
( a_-(\l) -(\frac{\l}{\mu})^{-2i+3})\kappa(\l)+\frac{1}{\kappa(\l)}\right)
\label{back}
\b
We next observe that the bracket (\ref{STS}) can be explicitely written as follows
\ai
\{f_{ij}(\l)\, , \, f_{ij}(\z)\}_{STS}&=&0\\ \label{bra}
\{f_{ii}(\l)\, ,  \, f_{ij}(\z)\}_{STS}&=&-2\frac{\l^2+\z^2}{\l^2-\z^2}f_{ii}(\l)f_{ij}(\z)+
4\frac{\l\z}{\l^2-\z^2}f_{ij}(\l)f_{ii}(\z)\\ \label{brb}
\{f_{ii}(\l)\, , \, f_{ji}(\z) \}_{STS}&=&2 \frac{\l^2+\z^2}{\l^2-\z^2}
f_{ii}(\l) f_{ji}(\z)-4\frac{\l\z}{\l^2-\z^2}f_{ji}(\l)f_{ii}(\z)\\ \label{brc}
\{f_{ij}(\l)\, , \, f_{ji}(\z)\}_{STS}&=&4\frac{\l\z}{\l^2-\z^2}
\left(f_{ii}(\l)f_{jj}(\z)-f_{jj}(\l)f_{ii}(\z)\right)\\ \label{brd}
\{f_{ii}(\l)\, , \, f_{jj}(\z) \}_{STS}
&=&4\frac{\l\z}{\l^2-\z^2}
\left(f_{ij}(\l)f_{ji}(\z)-f_{ji}(\l)f_{ij}(\z)\right)
\label{bre}\bj

It was shown in \cite{NBB} that taking into account the bracket (\ref{body}) for the one soliton solutions, the
Hamiltonians which generate the   $x^+$-- and $x^-$--- flows, are proportional to $\mu$ and $\mu^{-1}$, respectively. In view of this remark we define the "canonical" Hamiltonian 
\a
h(\l)&=&\frac{f_{11}(\l)-f_{22}(\l)}{f_{12}(\l)-f_{21}(\l)}
\label{hamc}
\b
This is justified by the fact that when $f(\l)$ is given by (\ref{mutandina}) one immediately gets
\a
h(\l)&=&
\frac{1}{2}\left(\frac{\l}{\mu}+\frac{\mu}{\l}\right)
\label{gen}\b
Due to that for the \dg\, elements (\ref{mutandina}), the expression 
(\ref{hamc}) is  a generating function of the $x^{\pm}$
flows.

Denote by $\tilde{\Gamma}$ the abelian subgroup of the loop group $\tilde{SL}(2)$ characterized by the restrictions $a_+(\l)=a_-(\l)=0$ in the
decomposition (\ref{Gauss}).
The element (\ref{mutandina}) belongs to $\tilde{\Gamma}$. Alternatively $\tilde{\Gamma}$ is described by the primary constraints \cite{vincoli}
\ai
c_1(\l)&=&f_{12}(\l)+f_{21}(\l)=0 \label{vuno} \\
c_2(\l)&=&f_{11}(\l)-f_{22}(\l)-\frac{\l^2+\mu^2}{2\l\mu}\left(f_{12}(\l)-f_{21}(\l)\right)=0  \label{vdue}
\bj

It is useful to use the Dirac's notation $\approx$ (weak identity):    functions on the loop group which coincide on  $\tilde{\Gamma}$ are said to be weakly equal. Taking into account (\ref{bra})--(\ref{bre}) we get the weak identities
\ai
\{ h(\l)\, , \, h(\z)\}_{STS}&\approx & 0\\ \label{apuno}
\{ h(\l)\, , \, c_1(\z)\}_{STS}&\approx & 2 \frac{(\l^2-\mu^2)\z}{\mu f_{12}(\l)(\l^2-\z^2)} \det \left (\begin{array}{cc}  \kappa(\l)& \kappa(\z)\\ 
\frac{\l^2-\mu^2}{\l\mu}f_{12}(\l)&  \frac{\z^2-\mu^2}{\z\mu}f_{12}(\z)
\end{array}\right ) \\ \label{apdue}
\{ h(\l)\, , \, c_2(\z)\}_{STS}&\approx & 0 \label{aptre}
\bj
Exploiting again (\ref{bra})--(\ref{bre}) and the constraints (\ref{vuno}),
(\ref{vdue}) we obtain
\ai
\{c_1(\l)\, , \, c_1(\z) \}_{STS}&=&0 \\ \label{pvinuno}
\{c_1(\l)\, , \, c_2(\z) \}_{STS}&\approx & 4\frac{(\l^2-\mu^2)\z}{(\l^2-\z^2)\mu}
\det \left (\begin{array}{cc}  \kappa(\l)& \kappa(\z)\\ 
\frac{\l^2-\mu^2}{\l\mu}f_{12}(\l)&  \frac{\z^2-\mu^2}{\z\mu}f_{12}(\z)
\end{array}\right ) \\  \label{pvidue}
\{c_2(\l)\, , \, c_2(\z) \}_{STS}&\approx &0  \label{pvintre} \bj
The above equations together with the primary constraints (\ref{vuno}), (\ref{vdue}) allow us to state that the brackets $\{h(\l)\, , \, c_j(\z)\}$ and $\{c_i(\l)\, , \, c_j(\z)\}$
only vanish  for the elements of $\tilde{\Gamma}$ for which the parameter $\kappa$ (\ref{Gauss}) does not depend on $\l$. We also note that imposing on
$\tilde{\Gamma}$ the condition of  $\l$--independence of $\kappa$  one recovers (\ref{mutandina}) provided that
\a
\kappa&=&e^{\frac{\var}{2}} \label{rel} \b
as a consequence of (\ref{campo}), (\ref{mutandina}) , (\ref{vuno}) and (\ref{vdue}). That is why we impose as a secondary constraint
\a
c_3(\l)=\frac{ d\, \kappa}{d\, \l}(\l)&=&0 \label{second} 
\b
Following the Dirac's algorithm we introduce the total Hamiltonian
\a
h_T(\l)&=&h(\l)+u^1(\l)c_1(\l)+u^2(\l)c_2(\l)
\label{tot}
\b
where $u^i(\l),\, i=1,2$ are Lagrange multipliers\footnote{More generally, instead of (\ref{tot}) one can consider the total Hamiltonian
$h_T(\l)=h(\l)+F^1[c_1](\l)+F^2[c_2](\l)$ where $F^i$ are linear functionals. However, in what follows we shall demostrate that (\ref{tot}) is sufficient to
get a relation between the Semenov--Tian--Shansky bracket and the bracket 
(\ref{body})}. Obviously
$h_T(\l)\approx h(\l)$.
Denote by $\F$ the subgroup of $\tilde{\Gamma}$ (\ref{vuno}), (\ref{vdue}) for which
$\kappa$ does not depend on $\l$. In what follows we shall use the
weak identity $\approx $ for quantities wich have equal values on
$\F$. The next step of the Dirac's procedure is to try to fix the values of the Lagrange multipliers $u^i(\l)$ by the conditions
\a
\{ h_T(\l) \, , c_i(\z)\}&\approx 0,\,\, i=1,2,3
\label{lag}
\b
The brackets $\{c_i(\l)\, , \, c_j(\l)\}$,
$i,j=1,2,3$ (\ref{vuno}), (\ref{vdue}), (\ref{second}) vanish on  $\F$. Due 
to that  $c_i$ are {\it first class constraints}. 
Taking into account the weak identities 
\a
\{h(\l)\, , \, \kappa(\z)\}&\approx & 0 \0\\
\{c_1(\l)\, , \, \kappa(\z)\}&\approx & 4f_{12}(\l)\kappa(\z) \0\\
\{c_2(\l)\, , \, \kappa(\z)\} & \approx & 0 \0\\
\{\kappa(\l)\, , \kappa(\z)\}& \approx & 0 
\label{app}
\b
we verify that the  Hamiltonian  $h$ (\ref{gen}) has a vanishing Poisson brackets with all the constraints fixing  $\F$. Therefore, the Lagrange multipliers in (\ref{tot}) can take arbitrary values. 
Imposing the  relation
\a
\{ h_T(\l)\, , \kappa(\z)\}_{STS}& \approx &\{h(\l)\, , \, \kappa\}
\label{crit}
\b
we only fix 
\a
u^1(\l)&=& \frac{1}{32} \left(\frac{\l^2-\mu^2}{\l\mu}\right)^2 (\kappa+ \kappa^{-1})
\b
as consequence of (\ref{body}) and (\ref{gen}).

We shall conclude by noting that since all the constraints are of the first class, our bracket (\ref{risultato}) arises after a suitable "gauge fixing" from the Semenov--Tian--Shansky bracket. 

\vspace{1truecm}

\centerline{\bf Acknowledgments}

R.P. would like to thank L. Bonora for discussions. 
The research of G. C. and R. P. were supported by Capes and CNPq respectively.
Both the foundations are gratefully acknowledged.

\end{document}